\documentclass{iopart} 
\usepackage{iopams}
\usepackage{graphicx} 
\newcommand{\rd}{\mathrm{d}} 
\newcommand{\fap}{\bar\mathbf{f}(\phi)}
\newcommand{\fn}{f_\mathrm{n}} 
\newcommand{\ft}{f_\mathrm{t}} 
\newcommand{\fnij}{f_{\mathrm{n},ij}}
\newcommand{\ftij}{f_{\mathrm{t},ij}} 
\newcommand{\fnap}{\bar{f}_\mathrm{n}(\phi)} 
\newcommand{\ftap}{\bar{f}_\mathrm{t}(\phi)} 
\newcommand{\Nab}{\mathcal{N}_{\alpha\beta}} 
\newcommand{\Tab}{\mathcal{T}_{\alpha\beta}} 
\newcommand{\taumax}{\tau_\mathrm{max}} 

\begin{document} 
\title{Bounds on the shear load of cohesionless granular matter} 
 
\author{Wouter G. Ellenbroek$^1$ and Jacco H. Snoeijer$^2$} 
\address{$^1$ Instituut--Lorentz, 
Universiteit Leiden, Postbus 9506, 2300 RA Leiden, The Netherlands}
\address{$^2$ School of Mathematics, University of Bristol, University Walk,
Bristol BS8 1TW, United Kingdom} 

\begin{abstract} 
We characterize the force state of shear-loaded granular matter by
relating the macroscopic stress to statistical properties of the force
network.  The purely repulsive nature of the interaction between
grains naturally provides an upper bound for the sustainable shear
stress, which we analyze using an optimization procedure inspired by
the so-called force network ensemble.  We establish a relation between
the maximum possible shear resistance and the friction coefficient
between individual grains, and find that anisotropies of the contact
network (or the fabric tensor) only have a subdominant effect. These
results can be considered the hyperstatic limit of the force
network ensemble and we discuss possible implications for real
systems.  Finally, we argue how force anisotropies can be related
quantitatively to experimental measurements of the effective elastic
constants.
\end{abstract} 
 
 
\section{Introduction} 

\begin{figure}[tbp] 
\includegraphics[width=8.5cm]{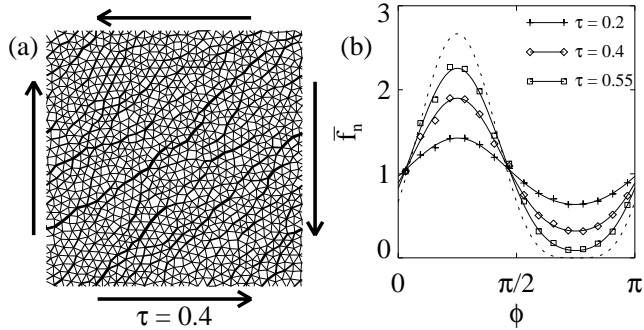} 
\centering 
\caption{(a) Force network obtained from a numerical simulation of a strongly
hyperstatic frictionless granular material (coordination number $z=5.75$)
subjected to a shear stress $\tau=\sigma_{xy}/\sigma_{xx}$, using the ``force
network ensemble''~\cite{forceensemble,worm}. The thicknesses of lines
represent the strength of the forces.
(b) Corresponding average contact force as a function of contact angle $\phi$
for this ensemble of force networks. Increasing the shear
stress yields a modulation of $\bar{f}(\phi)$ that is accurately described by
the form $1+2\tau \sin 2\phi + b_2 \cos 4\phi$. For large stress,
$\bar{f}(\phi)$ approaches the limiting curve predicted by
(\ref{limitingcurve}).} 
\label{fig1} 
\end{figure}

An assembly of cohesionless granular matter, in which there is no
attraction between grains, can only exist when held together by an
external pressure~\cite{bouchaudhouches}. The distribution of these
confining forces throughout the material is a complex process that
involves a highly inhomogeneous network of contact
forces~\cite{gm,radjai96,network,forceensemble}.
Force networks as shown in figure~\ref{fig1}a are typical for a broad
variety of amorphous systems like foams, colloids and emulsions, and
play a crucial role for understanding the macroscopic mechanical
properties~\cite{shearclement,brujic,bob}. 

A robust feature of these ``jammed'' materials is that they can sustain a
certain amount of shear stress before
failure~\cite{liunagel,daerr,nedderman,xuohern}.  There are many aspects that
influence this shear resistance or internal friction of a granular material. A
well known example is that wet sand can sustain much larger shear stresses than
dry sand, due to the presence of attractive liquid
bridges~\cite{naturekudrolli,albert}. The strength of the assembly is also
enhanced by increasing intergrain friction or roughness of the grains. However,
if one slowly increases the applied shear stress and follows the evolution of
contact forces and grain locations, one encounters very complex collective
phenomena.  Before the system yields as a whole, there are non-adiabatic
precursor events such as local rearrangements due to instability of subsets of
grains \cite{kabla,roux,oda}. This will induce changes in fabric and
coordination number, and it has remained a great challenge to understand how
this couples back to the stress
state~\cite{oda,radjairoux,dacruzPRE05,deboeuf}. The conventional tool to study
this problem is direct numerical simulation of the particle dynamics. While
this provides valuable information on the micromechanics of sheared granular
materials, it remains very difficult to distinguish the relative importance of
contact and force anisotropies: both are evolving simultaneously. 

Recently, a different strategy based on ``ensembles of force networks'' has
been proposed to address this problem~\cite{worm}. In this approach one
investigates the statistics of {\em all} possible  force network
configurations that are mechanically stable, for a {\em single} fixed
packing geometry of
grains~\cite{bouchaudhouches,forceensemble,unger,socolar}. This allows
explicit separation of the effects of forces and texture, e.g. by
studying the force ensembles for packs of different fabric and
coordination numbers. Consistent with direct simulations \cite{roux},
the ensemble showed that packings close to the minimum isostatic
coordination number can hardly support any stress whereas strongly
hyperstatic packings (much higher coordination numbers) can sustain
much more stress \cite{worm}. At the same time, the theory also
provides a very realistic description of the force anisotropy, force
fluctuations and response
function~\cite{forceensemble,worm,unger,socolar,mcnamara,ostojic}. 

In this paper we derive upper bounds for the shear load of
cohesionless granular media, for varying intergrain friction and fabric
anisotropy. The analysis is based on an intriguing observation made in
the force network ensemble for strongly hyperstatic packs. The most
elementary manifestation of force anisotropy is the modulation of the
average force $\bar{\bf f}(\phi)$, as a function of the contact
orientation $\phi$. From figure~\ref{fig1}b, obtained from a
two-dimensional frictionless system, it is clear that the force
anisotropy is limited by the requirement that the normal component
$\bar{f}_n(\phi)\ge 0$ for all $\phi$. This is due to the repulsive
nature of the contact forces which require all $f_{ij}\ge 0$. Indeed,
it was found that this simple criterion provides a very good
approximation of the maximum shear stress achieved in the force
network ensemble, in the limit of hyperstatic packs \cite{worm}. It also
gives a good prediction for the limiting form of $\bar{\bf f}(\phi)$ upon
approaching the maximum load.

The quantity $\bar{\bf f}(\phi)$ thus provides crucial information on
the force anisotropy and has recently been accessed experimentally for
the first time \cite{bob}. To generalize the arguments above to
frictional systems, we have to translate the physical constraints for
all individual contact forces, 
\begin{itemize}
\item normal forces are purely repulsive, i.e. $f_n \geq 0,$
\item tangential forces $f_t$ obey Coulomb's law of friction, 
$|f_t| \leq \mu\, f_n$,
\end{itemize}
to constraints for $\bar{\bf f}(\phi)$. Here, $\mu$ is the Coulomb
friction coefficient of the individual contacts. We show how this
yields nontrivial predictions for the maximum stress by finding the
extreme forms of $\bar{\bf f}(\phi)$. While these maxima can probably
not be reached in real systems, they describe the strongly hyperstatic
limit of the force network ensemble, making it a well-defined
analytical tool to investigate the influence of the micromechanical
parameters on the effective macroscopic friction. 

The optimization approach followed in the present paper is in the same
spirit as the force network ensemble, in the sense that it deals with
the question if a force network with given parameters can exist or
not, rather than with the actual evolution of a sheared
granular system. On the other hand, while force balance on each grain
is explicitly taken into account in the force network ensemble,
it plays no role in our analysis of $\fap$. This is allowed in the
strongly hyperstatic limit,
but not for nearly isostatic packs. In the latter case the force
networks are to a large extent determined by the conditions of force
balance~\cite{forceensemble,isostatic,isostatic2}.

The relevant macroscopic quantity is the deviatoric stress, defined as
$\tau=(\sigma_1-\sigma_2)/(\sigma_1+\sigma_2)$, where the $\sigma_i$
denote the principal values of the stress tensor. We work in the
coordinate frame where $\sigma_{xx}=\sigma_{yy}$, so we can express
the deviatoric stress as $\tau=\sigma_{xy}/\sigma_{xx}$.  For
cohesionless systems one may naively expect that the ultimate shear
stress is achieved when one principal direction becomes tensile, e.g.
$\sigma_2 =0$, which would lead to $\taumax=1$. By invoking realistic
structures of $\bar{\bf f}(\phi)$ for granular packs, however, we show
that the physics is in fact much more subtle, and that the real maximum
is typically much lower than unity.  Our main findings are that the
effect of friction between grains is only mild: a typical Coulomb
friction coefficient of $\mu=0.5$ increases $\taumax$ by only 16\% as
compared to the frictionless case. We also find that realistic
anisotropies in the contact fabric hardly increase $\taumax$ and thus
seem to play a subdominant role in real systems. 

\begin{figure}[tbp] 
\includegraphics[width=8.6cm]{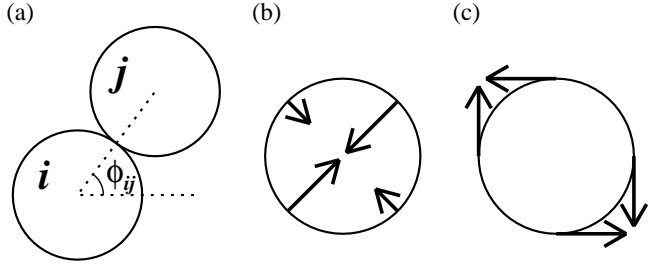} 
\centering 
\caption{(a) Definition of contact orientation $\phi_{ij}$. (b,c)
Illustration of the (average) bias of normal and tangential forces onto a particle 
due to the imposed shear stress.}
\label{fig2} 
\end{figure} 

The paper is organized as follows. We first define the quantities that
will be used to analyse the bounds, defined from the microscopic
structure of grains and contacts, in relation to the macroscopic shear
stress in section~\ref{sec.micromacro}. The third section addresses the
simplest case of isotropic, frictionless packings, and shows how the
granularity of the material affects the maximally supportable shear
stress. In sections~\ref{sec.frictional} and \ref{sec.anisotropy} we
explore the effects of intergrain friction and fabric anisotropies on
the maximum shear stress. We conclude in section~\ref{sec.discussion}, where we
argue how our approach can be applied to problems of anisotropic
elasticity.

\section{From contact forces to macroscopic stress}\label{sec.micromacro} 
 
We consider two-dimensional packings of discs, so that the orientations
of the contacts between particles $i$ and $j$ can be characterized by
the angle (figure~\ref{fig2}a)
\begin{equation} 
\label{defphi} 
\phi_{ij}=\arccos\left(\frac{x_j-x_i}{|\mathbf{r}_j-\mathbf{r}_i|}\right). 
\end{equation} 
Here $\mathbf{r}_i$ denotes the position vector of particle $i$ and
$x_i$ its $x$-coordinate. The key quantity that we will use to
determine the maximum shear load is the average force $\fap$
carried by all contacts of orientation $\phi$. Since bond directions
have no polarity, the angle only assumes values $0\leq\phi<\pi$, and
the period of $\fap$ is $\pi$. One can relate $\fap$ to the stress
tensor $\sigma$, using~\cite{goldhirsch}

\begin{equation}\label{stresss} 
\sigma_{\alpha\beta}=\frac{1}{V}\sum_{\{ij\}}(\mathbf{f}_{ij})_\alpha(\mathbf{r}_{ij})_\beta
\equiv \frac{N_c}{V} \, \overline{\, \mathbf{f}_\alpha \mathbf{r}_\beta}
~. 
\end{equation} 
Here $\alpha,\beta$ label coordinate axes, $N_c$ is the number of contacts in 
the (two-dimensional) volume $V$, $\mathbf{r}_{ij}=\mathbf{r}_j-\mathbf{r}_i$, and $\mathbf{f}_{ij}$ is the force exerted on particle $j$ by particle $i$. The bar indicates average over all forces in $V$. We decompose the force vector in a normal component, $\fnij$, and tangential components, $\ftij$, as
\begin{eqnarray} 
(\mathbf{f}_{ij})_x&=&\fnij\cos\phi_{ij}-\ftij\sin\phi_{ij} \label{fijx}\\ 
(\mathbf{f}_{ij})_y&=&\fnij\sin\phi_{ij}+\ftij\cos\phi_{ij}~. \label{fijy} 
\end{eqnarray} 
The sign conventions are such that repulsive forces have positive
$\fnij$, while the tangential component is positive when pointing
``counterclockwise'' with respect to particle $i$ (see
figure~\ref{fig2}). 

For large enough packings we can express the stress tensor in a
statistical form, evaluating the average $\overline{\,
\mathbf{f}_\alpha \mathbf{r}_\beta}$ from the probability to find a
contact with force $\mathbf{f}_{ij}$ and centre-to-centre vector
$\mathbf{r}_{ij}$. In terms of normal and tangential components,
and with the observation that the forces are uncorrelated to the
interparticle distance $| \mathbf{r}_{ij}|$~\cite{worm,radjaiPRL98}, this
involves the joint probability $P(\fn,\ft,\phi)$. We can explicitly
factorize the contact angle probability $\Phi(\phi)$ to write 
$$ 
P(\fn,\ft,\phi)=\Phi(\phi)P_\phi(\fn,\ft). 
$$ 
The distribution $P_\phi(\fn,\ft)$ has been introduced recently~\cite{worm} and is properly normalized to unity. Hence, the probabilistic form of the stress tensor reads 

\begin{eqnarray} 
\sigma_{\alpha\beta} &=& \frac{{\bar r} N_c}{V} \, 
\int_0^{\pi} \rd\phi \, \Phi(\phi) \int_0^\infty \rd f_\mathrm{n} 
\int_{-\infty}^\infty \rd f_\mathrm{t} \nonumber\\ 
&&\times P_\phi(f_\mathrm{n},f_\mathrm{t})\left[f_\mathrm{n}\Nab+f_\mathrm{t}\Tab\right] 
\nonumber \\ 
&=& 
\frac{{\bar r} N_c}{V} \, \int_0^{\pi} \mathrm{d}\phi \, \Phi(\phi)\, 
\left[\fnap\Nab+\ftap\Tab\right],\label{stresstensor} 
\end{eqnarray} 
where $\bar{r}$ denotes the average interparticle distance. The projection factors have been collected in tensors $\Nab$ and $\Tab$, written in matrix notation as 
\begin{eqnarray} 
\Nab&=& \left( 
\begin {array}{cc} 
\cos^2 \phi&\cos \phi \sin \phi \\ 
\cos\phi \sin \phi & \sin^2 \phi 
\end {array}\right)\\ 
\Tab&=&\left( 
\begin {array}{cc} 
-\cos \phi \sin \phi &-\sin^2 \phi \\ 
\cos^2 \phi & \cos \phi \sin \phi 
\end {array} 
\right)~. 
\end{eqnarray} 
In the remainder, we will set the prefactor ${\bar r} N_c/V =1$ in (\ref{stresstensor}). 

The above analysis allows computing the stress from $\fap$. To relate
the force anisotropies to the shear stress, a common trick is to
expand $\fap$ in a Fourier series~\cite{dacruzPRE05,worm,radjaiPRL98}
\begin{eqnarray} 
\fnap&=&\sum_{k=1}^N \,a_k\sin 2k\phi + \sum_{k=0}^N \, b_k \cos 2k\phi \label{expfn}\\ 
\ftap&=&\sum_{k=1}^N \,c_k\sin 2k\phi + \sum_{k=0}^N \, d_k \cos 2k\phi \label{expft}~. 
\end{eqnarray} 
Because we are working in the frame where $\sigma_{xx}=\sigma_{yy}$, the
principal axes of stress point in the directions $(1,1)$ and $(1,-1)$.
These directions must then be lines of mirror symmetry, as is illustrated
in figure~\ref{fig2}, which makes that $a_k=d_k=0$ for even $k$ and $b_k=c_k=0$
for odd $k$.  

For the moment, until the last section of the paper, we will consider the case
where the fabric is isotropic so $\Phi(\phi)=1/\pi$. In that case, inserting
equations (\ref{expfn},\ref{expft}) in (\ref{stresstensor}) and integrating
yields 
\begin{eqnarray} 
\sigma_{xx}&=&b_0/2\\ \label{sxxiso} 
\sigma_{yy}&=&b_0/2\\ \label{syyiso} 
\sigma_{xy}&=&(a_1+d_1)/4 \label{sxyiso}~. 
\end{eqnarray} 
All higher order terms in the expansion yield zero upon integration. Our main interest is the deviatoric stress, so we are free to choose the pressure scale as $\sigma_{xx}=\sigma_{yy}=1/2$, so that $\bar{f}=b_0=1$ and 
\begin{equation} 
\label{tauforward} 
\tau=\frac{a_1+d_1}{2}~. 
\end{equation} 
This relation reveals how an applied shear stress can be sustained
through anisotropies in both the normal and frictional forces, via
$a_1$ and $d_1$ respectively~\cite{radjaiPRL98}. The strategy will be
to explore the physical limitations of $a_1$ and $d_1$, which will
provide a bound on $\tau$. 

Note that due to the linearity of (\ref{stresss}), the stress only couples to the first moment of the force distributions. This means that the stress does not depend on details of the probability
$P(|{\bf f}|)$~\cite{radjai96,network,forceensemble,brujic,bob}.
Also, the stress tensor contains no information on ``force
chains'', i.e.\ the tendency of large forces the align in a correlated way.
(\ref{stresss}) does not invoke products of different ${\bf f}_{ij}$ so
that spatial force-force correlations do not come into play.

\section{Frictionless packings}
\label{sec.frictionless}
We start out with packings of frictionless particles, for which the problem of
the maximum possible deviatoric stress is relatively straightforward. In this
case a bound on $\tau$ emerges from the purely repulsive nature of the forces:
all contacts have $\fnij\ge0$, so certainly the averages should obey 

\begin{equation}\label{repulsive}
\fnap\ge0~,
\end{equation} 
for all values of $\phi$. This condition obviously forms a serious
restriction on the amplitude of the force anisotropy, and we show how
this yields an upper limit on $\tau$. We show that this maximum
value $\taumax$ depends on the number $N$ of terms included in the
Fourier series of (\ref{expfn}): even though the higher order terms 
do not contribute to the stress, they enable $\fnij$ to reach more extreme shapes. 
We first explore the case $N=2$, which is the relevant case for real
systems (see also \ref{appcoord}). Then we discuss the problem for
arbitraty $N$, which provides some additional insights.

\subsection{Realistic $\fnap$}\label{subsecfrictionless}

For the frictionless case we have from (\ref{tauforward}) 
\begin{equation} 
\label{tauback1} 
a_1=2\tau,\qquad d_1=0, 
\end{equation} 
so that (\ref{expfn}) becomes
\begin{equation} 
\label{fn1} 
\fnap=1+2\tau\sin 2\phi + b_2\cos 4\phi +\cdots 
\end{equation} 
Let us first consider the simplest case, in which we truncate after
the lowest anistropic term, i.e.\ $\fnap=1+2\tau\sin2\phi$, so that
$\tau$ is the only free parameter. For positive $\tau$ this function
has a minimum in $\phi=3\pi/4$, which touches $\fnap=0$ for $2\tau=1$.
Hence, $\taumax=1/2$. In the same way we could derive $\tau_\mathrm{min}=-1/2$. From now on we will only consider positive $\tau$, without loss of generality. 
 
From the numerical result of figure~\ref{fig1}b it is clear that the
modulation does not stay symmetric around $\fnap=1$ for large values
of $\tau$. This indicates a significant contribution of the type $\cos
4\phi$.
This is the highest order term that we can expect to be relevant in
granular matter, because steric exclusion between the grains sets a lower
limit to the width of peaks in $\fnap$. This is explained in more detail
in \ref{appcoord}.

We thus focus on the case $N=2$ in the expansion~(\ref{expfn}). The
optimization problem involves two free parameters, $\tau$ and $b_2$.
We are free to vary $b_2$ in such a way as to facilitate a maximum
$\tau$, under the constraint that $\fnap\ge0$ for all $\phi$. We
furthermore demand that $\fnap$ evolves monotonically between the
principal directions at $\phi=\pi/4$ and $\phi=3\pi/4$. This implies
that the minimum of $\fnap$ should stay at $\phi=3\pi/4$, as it is the
case in the numerics of figure~\ref{fig1}b. Physically, this means that the first contacts to break are those oriented in the direction in which the material is stretched. While the expansion ensures that there is an extremum in $\phi=3\pi/4$, this extremum only remains a minimum if 
$$ 
\left.\frac{\partial^2}{\partial\phi^2}\fnap\right|_{\phi=\frac{3\pi}{4}}= 
8\tau+16b_2\ge0~. 
$$ 
The value of $\fnap$ in this minimum must satisfy 
$$ 
\bar{f}_\mathrm{n}(3\pi/4)=1-2\tau-b_2\ge0~. 
$$ 
This defines a linear program with parameters $\tau$ and $b_2$, two inequalities,
and the objective to maximize $\tau$. The solution is easily found to
be~\footnote{Actually, because for these values of the
parameters the second derivative vanishes in $\phi=3\pi/4$, we
have to check that the fourth derivative is positive to be sure
that the extremum is a minimum, which is indeed the case.}
\begin{eqnarray} 
\taumax&=&\frac{2}{3}\label{tau2}\\ 
b_2&=&-\frac{1}{3}~.\label{b22} 
\end{eqnarray} 
figure~\ref{fig1}b illustrates the relevance of this bound for strongly
hyperstatic packings: the numerical $\fnap$ (taken from~\cite{worm}) indeed approaches
the limiting form (dashed curve)

\begin{equation}\label{limitingcurve}
\fnap=1+\frac{2}{3}\sin 2\phi - \frac{1}{3} \cos 4\phi~.
\end{equation}
So indeed, the system is able to organize the forces in such a manner as to optimize
the sustained shear stress. 

\subsection{The limit $N\rightarrow \infty$ and tensile stresses} 
\begin{figure}[tbp] 
\includegraphics[width=8.6cm]{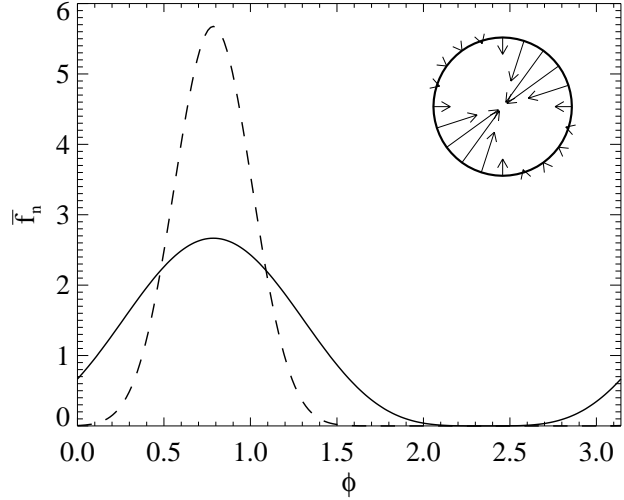} 
\centering 
\caption{The optimized $\fnap$ for $N=2$ (solid), $N=10$ (dashed). 
The figure in the corner illustrates the corresponding average force exerted
on a particle for various contact orientations ($N=2$).} 
\label{fig3} 
\end{figure} 
Although Fourier terms with $k\ge 3$ do not play a role in granular
systems, it is insightful to study the general case where we truncate
the series at arbitrary order $N$. The expansion now contains $N$ free
coefficients that we need to fix, in order to optimize $\tau$. In the
case $N=2$, we invoked the condition $\partial^2 \bar{f}_{\rm
n}/\partial \phi^2\ge0$ at $\phi=3\pi/4$ to ensure that the minor
principal axis remains a minimum of $\fnap$, and taking $\partial^2
\bar{f}_{\rm n}/\partial \phi^2=0$ turned out to maximize $\tau$.  A
nonzero $k=3$ term allows to also fix $\partial^4 \bar{f}_{\rm
n}/\partial \phi^4$ in such a way that $\tau$ becomes even larger.
The upshot of adding more terms is that we can make $\fnap$ as flat
(and as close to zero) as possible around $\phi=3\pi/4$,
where the contribution to the overlap with $\sin2\phi$ is negative.
Every time we have to verify that the first nonzero Taylor coefficient when
expanding around $\phi=3\pi/4$ is positive, so that we are indeed dealing
with a minimum.

The general scheme is thus that adding the term of order $k$ generates an
additional condition that $\partial^{(2k-2)} \bar{f}_{\rm n}/\partial
\phi^{(2k-2)}=0$. For general $N$, this yields the following set of linear
equations:

\begin{equation}\label{arbitrary}
\left. \frac{\partial^{2l}}{\partial \phi^{2l}}\, \fnap \, 
\right|_{\phi=\frac{3\pi}{4}}=0 
\quad {\rm for} \,\,\, {\rm all} \quad l=0,2,\cdots, N-1~. 
\end{equation}
In \ref{appvandermonde} we show that this linear problem for the Fourier coefficients can be inverted analytically, yielding a remarkably simple result for the maximum $\tau$, namely

\begin{equation}\label{tauN}
\taumax(N) = \frac{N}{N+1}~.
\end{equation}

Interestingly, this reveals an ultimate (mathematical) maximum shear
stress $\taumax=1$. We plotted the optimized $\fnap$ for various
values of $N$ (figure~\ref{fig3}): clearly, $\fnap$ evolves towards the
extreme case of a Dirac $\delta$-peak in the limit $N\rightarrow
\infty$. As already mentioned in the introduction, this condition of
$\tau=1$ precisely corresponds to the point where the minor principal
axis becomes tensile. We thus conclude that, due to the finite width
of $\fnap$, the maximum stress for a frictionless packing lies well
below the point where the global stress unavoidably develops a tensile
direction.
We have argued in \ref{appcoord} that this finite width is in fact due
to steric exclusion between neighbouring grains. This illustrates how
the discrete nature of the assembly has an important effect on global
properties.

\section{Frictional packings: $\taumax(\mu)$}\label{sec.frictional}
 
The presence of frictional forces provides an additional degree of freedom
to develop anisotropic stresses: Equation~(\ref{tauforward}) shows that the
total deviatoric is the sum of the (lowest order) anistropies of normal
and tangential forces. It is clear that this will enhance the ability to
sustain a large external load. However, there is again a bound on the
force anisotropies, now due to Coulomb's law for individual contacts,
i.e.\ $|f_{{\rm t},ij}|\le \mu f_{{\rm n},ij}$, where $\mu$ is the microscopic
Coulomb friction coefficient. Since this condition should hold
for any pair of grains, it certainly holds for the averages:

\begin{equation}\label{coulombaverage}
|\ftap | \, \le \, \mu \fnap~.
\end{equation}
This condition is illustrated in figure~\ref{fig4}. In this section we
derive how $\taumax$ depends on the value of $\mu$, again using the
Fourier expansions of $\fnap$ and $\ftap$ up to $N=2$.

\subsection{The optimization problem}\label{subsec.fricopt}
From (\ref{tauforward}) we can express
\begin{equation}
a_1=2\tau-d_1~,
\end{equation}
so that 
\begin{eqnarray} 
\label{fn2} 
\fnap&=&1+(2\tau-d_1) \sin 2\phi + b_2\cos 4\phi ~, \\
\ftap&=& d_1 \cos 2\phi + c_2\sin 4\phi ~.
\end{eqnarray} 
Let us now take a value of $\tau$ slightly above the frictionless limit
$2/3$. The prefactor in front of the $\sin 2\phi$ term can now be lowered
due to $d_1$, i.e.\ due to the presence of friction. If we put $\tau$
far above $2/3$, however, one requires a relatively large $d_1$. The
value of $d_1$ is bounded by the condition of (\ref{coulombaverage}),
so that not all $\tau$ can be reached. 

\begin{figure}[tbp] 
\includegraphics[width=8.6cm]{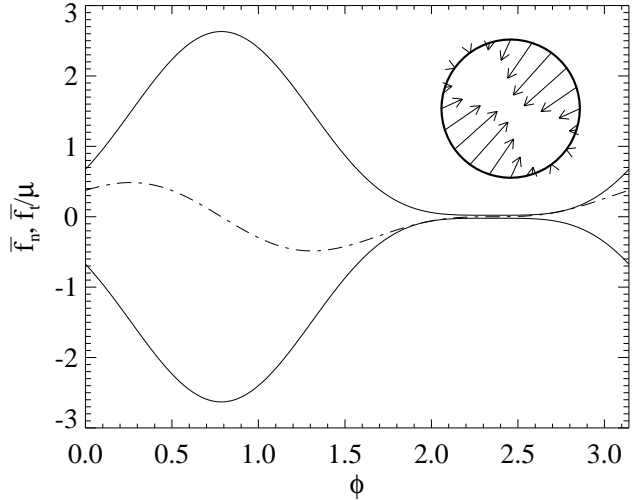} 
\centering 
\caption{The frictional optimization problem for $\mu=0.28$.
The solid lines represent $\pm\fnap$, the dash-dotted line $\ftap/\mu$,
which has to satisfy (\ref{coulombaverage}). 
The figure in the corner illustrates the average force exerted on a frictional
particle for various contact orientations. Note that these forces now have
tangential components.} 
\label{fig4} 
\end{figure} 
To determine the maximum value of $\tau$ as a function of $\mu$, we have
to specify acceptable values of the higher order coefficients $b_2$
and $c_2$, which do not contribute to the stress tensor. As we did in the
frictionless case, we demand that the function $\fnap$ evolves
monotonically between major and minor directions, so that it has only
one maximum (in the major direction) and only one minimum (in the minor
direction). 
A similar requirement is imposed on $\ftap$: the average tangential
force only changes sign along the principal directions, see e.g.
figure~\ref{fig4}. If this were not the case, the tangential forces
would swap from clockwise to counterclockwise and back in between the
major and minor directions. Such a spontaneous symmetry breaking would
introduce a very nongeneric organization of forces within the packing. 
These conditions put bounds on the second order coefficients,
\begin{eqnarray}\label{boundb2}
-\frac{2\tau-d_1}{4} \le &b_2& \le \frac{2\tau-d_1}{4}~, \\
-\frac{d_1}{2}\le &c_2& \le \frac{d_1}{2}~.
\label{boundc2}
\end{eqnarray}

\begin{figure}[tbp] 
\includegraphics[width=8.6cm]{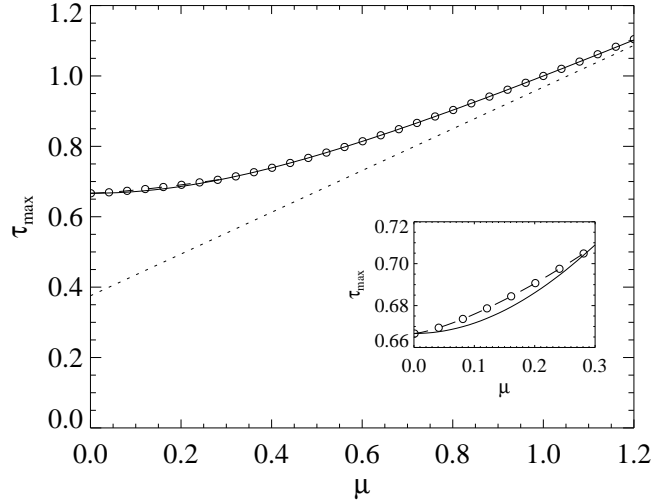} 
\centering 
\caption{The optimized $\tau(\mu)$ for frictional packings with
isotropic fabric. The circles are numerical data. The solid line is
equation (\ref{taufricresult}). The dashed line is the analytic result for
$\mu\leq 1-\frac12\sqrt2\approx0.29$. The dotted line is the asymptote $\tau \propto
0.593 \mu$ as $\mu\to\infty$. The inset shows an enlargement of the small
$\mu$ region.} 
\label{fig5} 
\end{figure}

We have numerically solved this optimization problem by varying all
possible values of the parameters $\tau, d_1,b_2,c_2$, within the ranges
imposed by equations~(\ref{coulombaverage},\ref{boundb2},\ref{boundc2}). The
results are shown in figure~\ref{fig5}. Surprisingly, the dependence on $\mu$ is
relatively weak. In the paragraph below we obtain the analytical result,

$$
\taumax=\frac{1+\sqrt{1+3\mu^2}}{3}~,
$$
which is derived for $1-\frac12\sqrt2\leq\mu\leq 1$.
However, figure~\ref{fig5} shows that this is a very good approximation
outside this range as well. 

\subsection{Analytical solution of $\tau(\mu)$}
The set of parameters $\tau,d_1,b_2,c_2$ that corresponds to the
maximum value of $\tau$ obviously corresponds to functions $\ftap$ 
and $\mu\fnap$ that are tangent in at least one point. We will now
derive the optimal set of parameters for the case there is a
tangent point in the interval $\frac{7\pi}8\leq\phi\leq\pi$, and
then determine for what range of $\mu$ this is the case.

In the interval $[\frac{7\pi}8,\pi]$, the $\cos2\phi$ in $\ftap$ is
positive and the $\sin4\phi$ is negative. We want to have a $d_1$
which is large (to facilitate larger $\tau$), and at the same time
a $\ftap$ which is as close to zero as possible (to stay away from 
violating (\ref{coulombaverage})). Therefore, the parameter
$c_2$ should have its maximum value $c_2=d_1/2$, as dictated by
(\ref{boundc2}). Similarly, we want $\fnap$ to be as large as
possible, and hence because $\cos4\phi$ is positive in the
considered interval, $b_2$ should also be maximal. There are two
upper bounds on $b_2$, given by (\ref{boundb2}) and $\fnap\geq0$, the
latter of which turns out to be the most restrictive one. This gives
$$
b_2=1+d_1-2\tau~.
$$
Having eliminated $b_2$ and $c_2$, we have thus reduced the optimization
problem to 2 parameters, namely $\tau$ and $d_1$. To find the
parameters that maximize $\tau$, we take the equality sign in
(\ref{coulombaverage}) and demand that solutions are tangent
points. For arbitrary values of the parameters, there is a tangent point
in $\phi_1=3\pi/4$ and two normal intersection points in $\phi_2$ and
$\phi_3$. The intersection points should coincide to turn into a tangent
point. Equating $\phi_2=\phi_3$ gives a relation between the parameters
which, after some lengthy but elementary algebra, can be written as
$$
\tau=\frac12d_1+\frac13+\frac{1}{6\mu}\sqrt{4\mu^2-3d_1^2}~.
$$
Maximizing this expression for $\tau(\mu,d_1)$
with respect to $d_1$ yields that the optimum is obtained for
$$
d_1=\frac{2\mu^2}{\sqrt{1+3\mu^2}}~,
$$
so that
\begin{equation}
\label{taufricresult}
\taumax(\mu)=\frac{1+\sqrt{1+3\mu^2}}{3}~.
\end{equation}
Inserting these into the original equations gives the value of $\phi$
where the curves touch,
\begin{equation}
\tan2\phi=\frac{\sqrt{1+3\mu^2}-2}{3\mu}~,
\end{equation}
which was demanded to be in the interval
$[\frac{7\pi}8,\pi]$. This requirement is met for values of $\mu$ satisfying
\begin{equation}
(0.29\approx)~1-\frac12\sqrt2\leq\mu\leq1~.
\end{equation}
This validity is indeed found numerically in figure~\ref{fig5}. Note
that this range of Coulomb coefficients is the most relevant for real
granular materials \cite{coulombtable}.

For smaller $\mu$ the tangent point is below $\phi=\frac{7\pi}8$,
where the $\cos4\phi$ is negative, so $b_2$ should now be as small as
possible: $b_2=-(2\tau-d)/4$. The resulting quartic equations can be
solved using computer algebra, yielding a lenghty expression (not
shown), which is plotted as the dashed part of the curve
in figure~\ref{fig5}, and which coincides with the numerical data. 

For $\mu>1$ the above analysis gives a tangent point in the interval
$[0,\pi/12]$, so that
the considerations that allowed us to fix $b_2$ and $c_2$ are no longer
valid and we only have the numerical result of figure~\ref{fig5}.
When the Coulomb coefficient becomes very large, the sustainable
shear stress is mostly due to the tangential forces. Because the right
hand side of (\ref{coulombaverage}) grows linearly with $\mu$, the
values of $d_1$ and $c_2$ can also scale as $\mu$ when $\mu\gg1$. This
leads to an to an asymptotic behaviour of $\taumax(\mu)$ which is
linear in $\mu$. We numerically determined the parameters
$a_1,b_2,c_2/\mu,d_1/\mu$ for the asymptotic $\fnap$ and $\ftap/\mu$ by
optimizing for $d_1/\mu$ instead of $\tau$. The resulting asymptote is
$\tau=0.376+0.593\mu$, which plotted as the dotted line in figure~\ref{fig5}.

\section{Fabric anisotropy}\label{sec.anisotropy}
\begin{figure}[tbp]
\includegraphics[width=8.6cm]{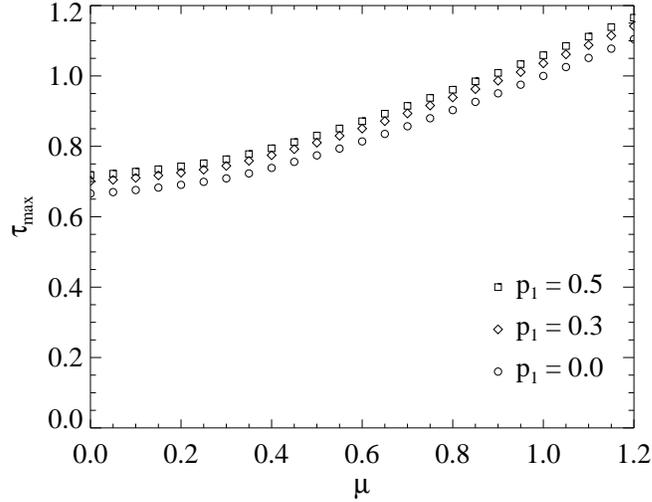}
\centering
\caption{Maximum shear stress $\taumax$ as a function of friction
coefficient $\mu$ for $p_1=0,0.3,0.5$.}
\label{fig6}
\end{figure} 
We now analyze the effect of a fabric anisotropy on the maximum shear
stress. The structure of the contact network can be characterized
using the fabric tensor $F_{\alpha\beta}=\overline{\,
\mathbf{r}_\alpha \mathbf{r}_\beta/r^2}$, and a fabric anisotropy
shows up as a difference between the principal values of this tensor,
$F_1-F_2$. Analogous to the stress tensor, this difference is solely
determined by the lowest order coefficient of the Fourier expansion of
$\Phi(\phi)$, the contact angle distribution, which we therefore
approximate as~\footnote{The maximum shear stress is achieved when contact
and force anisotropy are ``in phase'', so that we do not introduce a
phase shift in the $\sin 2\phi$ term of (\ref{contactfourier}).}
\begin{equation}\label{contactfourier}
\Phi(\phi)=\frac{1}{\pi}(1+p_1\sin2\phi)~.
\end{equation}
One easily shows that $p_1=2(F_1-F_2)$. The strategy is to again find the
maximum possible value for $\tau$, but now for $p_1 \neq 0$. 

Let us note that this truncated form for $\Phi(\phi)$ is a good
approximation for systems with a simple shear deformation
history~\cite{radjaiPRL98,alonsoPRE2005}, although more complex forms are
encountered e.g. for packings formed under gravity \cite{atmanfabric}. As was
demonstrated by Troadec {\em et al.}~\cite{troadec}, the values for $p_1$
are bounded by the effect of steric exclusion between neighbouring contacts
(in the same way as steric exclusion bounds the force anisotropy,
see also \ref{appcoord}). Indeed, numerical simulations show that
$p_1<0.3$, while typically $p_1$ is of the order
$0.1$~\cite{radjaiPRL98,alonsoPRE2005}. 

We thus insert the form (\ref{contactfourier}) in the expression for the stress tensor (\ref{stresstensor}), yielding
\begin{eqnarray}
\sigma_{xx}&=&b_0/2+p_1a_1/4\label{anisxx}\\
\sigma_{yy}&=&b_0/2+p_1a_1/4\label{anisyy}\\
\sigma_{xy}&=&\frac{1}{4}(a_1+d_1)+\frac{p_1}{8}(2b_0+c_2-b_2)\label{anisxy}~.
\end{eqnarray}
This time the second order coefficients {\em do} affect the value of the
deviatoric stress directly. The idea of the optimization procedure,
however, remains the same as in the isotropic case. In this section we
present the results of the optimization. The frictionless case, which has
an analytic solution, is discussed in more detail in \ref{appanis}.

We again eliminate $b_0$ by setting $\sigma_{xx}=1/2$ in (\ref{anisxx}).
From (\ref{anisxy}) we then have
\begin{equation}
\label{tauanisfric}
\tau=\frac{\sigma_{xy}}{\sigma_{xx}}=\frac{a_1(2-p_1^2)+2d_1+p_1(2-b_2+c_2)}{4}~.
\end{equation}
Working out the optimization for the frictionless case ($c_2=d_1=0$),
we get the analytic result
\begin{equation}
\label{taumaxanis}
\taumax=\frac{7p_1+8}{8p_1+12}~,
\end{equation}
as is shown in \ref{appanis}.
The inclusion of fabric anisotropy hence leads to a small increase of
the maximum deviatoric stress; for e.g.\ $p_1=0.3$ we get a 5\%
increase. See also the numerical data points on the vertical axis of
figure~\ref{fig6}.

For the complete case, with anisotropic fabric and finite friction
coefficient, we have used the same numerical optimization procedure as in
section~\ref{subsec.fricopt}. The result is shown for $p_1=0.3,0.5$ in
figure~\ref{fig6}. From this figure one can see that the combined effect
of friction and anisotropy can roughly be seen as an addition of their
individual effects.

\section{Discussion}\label{sec.discussion}

The repulsive-only interaction between dry grains causes the shear stress
sustained by granular materials to be bounded. We have derived these upper
bounds, $\taumax$, by finding the extreme shapes of the angle-resolved
average force, $\bar{\bf f}(\phi)$, that are consistent with Coulomb's
friction law for individual grains. In the spirit of the Force Network
Ensemble, the key question addressed in the present paper is ``can such a
force network exist?'', and we have argued that our analysis represents
the strongly hyperstatic limit of the ensemble. 

The yielding process in real systems is much more involved and cannot
be fully described by the existence criteria used in the paper.
However, the analysis does allow to investigate the influence of the
micromechanical parameters on the maximum shear stress, in such a way
that different aspects can be disentangled. This provides useful
complementary information to simulation methods, for which the packing
texture cannot be controlled during loading. For example, we have
found that an anisotropic fabric is not needed to sustain a large
shear stress, and in fact adding fabric anisotropy hardly increases
the maximum stress. This suggests that shear-induced textures observed
in numerical simulations~\cite{radjaiPRL98,alonsoPRE2005} play a
relatively passive role in the stress balance. 

Secondly, we analyzed the effect of the microscopic friction coefficient $\mu$,
and found that $\taumax$ does not increase rapidly with $\mu$ in the physically
relevant regime (figure~\ref{fig5}). One has to be very careful, however, to
interpret this result. The presence of friction {\em does} have an important
influence on the coordination number, which in turn affects the shear
resistance: our analysis breaks down close to the isostatic limit,
at which the force balance equations become dominant and
$\taumax$ drops to zero~\cite{worm}.
Consistent with our analysis, the internal friction
measured in simulations of quasi-static shear
flows~\cite{dacruzPRE05,tamascondmat} displays only a mild increase is
observed when $\mu > 0.4$. On the other hand, $\tau$ was found to decrease
rapidly as $\mu \rightarrow 0$, which we believe is because the system
approaches the isostatic limit~\footnote{It should be noted that it is
not obvious to determine
how far from isostaticity (marginal stability) a system is.  Two-dimensional
systems of frictionless disks in principle become isostatic at $z=3$, but when
the microscopic friction coefficient $\mu$ is small, many contacts are fully
mobilized (i.e.\ the frictional force has its maximum value allowed by the
Coulomb criterion) \cite{kostya}. These contacts
contribute less to the number of degrees of freedom in the force network, since
the tangential force component is fixed by the normal force component
\cite{bouchaudhouches,kostya}. Therefore, these systems are less
hyperstatic than one would think on the basis of the coordination number.}.

Our main conclusion is thus that strongly hyperstatic packings of grains can
support an amount of shear stress that is not very sensitive to friction or
fabric anisotropy. For systems close to the isostatic limit, however, the
dependence of the coordination number on any of these parameters has a
more dramatic effect.
 
Another interesting perspective is that one can use an expansion of
$\bar{\bf f}(\phi)$, as presented in this paper, to
estimate the effective elastic moduli of the system, denoted by
$C_{\alpha\beta\gamma\delta}$. It has been shown experimentally that these
become highly anisotropic when a system is
pre-sheared~\cite{shearclement,atmanbrunet}.  The response to a localized
vertical force on the free surface of a granular bed was found not to propagate
along the vertical, but along a direction that is tilted towards the major
stress axis. This indicates a stiffening of contacts along the major axis that
are responsible for the anisotropic elasticity.  By fitting the experimentally
measured stress profiles Atman {\em et al.}~\cite{atmanbrunet} obtained the
ratio $C_{1111}/C_{2222}\approx 0.67$ of the Young's moduli in the minor (1)
and major (2) directions. This finding was remarkably insensitive to frictional
properties and roughness of the grains.

The stiffening of contacts can be explained through the nonlinear
interaction between particles, which for frictionless Hertzian
contacts~\cite{Johnson} in three dimensions is $k_{ij} \propto
f_{ij}^{1/3}$, where $k_{ij}$ is the effective spring-constant of the
contact. Assuming that the displacements of the particles are
affine --- this is reasonable for the strongly hyperstatic packings considered 
in this paper~\cite{linresp} --- one can estimate the elastic moduli
as~\cite{elasticwalton,elasticbath}~%
\footnote{The use of $(\overline{f})^{1/3}$ instead of
$\overline{f^{1/3}}$ introduces only a small error since these functions
differ mainly
around the minor principal axis where the integrand is small anyway.}
\begin{eqnarray}
C_{\alpha\beta\gamma\delta} &\propto& \int_0^\pi d\phi \, \Phi(\phi) \bar{k}(\phi)
n_\alpha n_\beta n_\gamma n_\delta \nonumber \\
&\propto& \int_0^\pi d\phi \, \Phi(\phi) \left[ \bar{f}(\phi) \right]^{1/3} 
n_\alpha n_\beta n_\gamma n_\delta~.
\end{eqnarray}
This allows to systematically explore the effect of stress induced
anisotropy in forces or fabric, by again using the Fourier expansions
of $\bar{f}(\phi)$ and $\Phi(\phi)$. For the simplest case of
isotropic frictionless contacts, one already finds that $C_{1111}/C_{2222} =
1-8\tau/9 +{\cal O}(\tau^2)$, so that the experimentally observed
ratio of about 0.67 is easily achieved for realistic values of $\tau$. 
 
\ack
We thank Martin van Hecke, Thijs Vlugt, Julien Tailleur,
Philippe Brunet, Kostya Shundyak,
and Wim van Saarloos for numerous discussions. WGE acknowledges
financial support from the physics foundation FOM, JHS from a Marie
Curie European Fellowship FP6 (MEIF-CT-2006-025104). 
 
\appendix

\section{Steric exclusion and the coordination number}\label{appcoord}
We show in section \ref{sec.frictionless} that optimizing $\taumax$ is equivalent to making the maximum
of $\bar{\bf f}(\phi)$ around the major principal axis as sharp as
possible.
It has been reported from
numerical simulations of realistic systems that in practice this maximum is not
very sharp and that typically the highest Fourier contribution to $\bar{\bf
f}(\phi)$ is of order $\cos 4\phi$~\cite{worm,radjaiPRL98}
(see also figure~\ref{fig1}b). This
can be related to the coordination number as follows.  Suppose we wish to
maximize the forces around the major principal axis, $\phi=\pi/4$. On a typical
grain this can be achieved by putting all stress on the contact closest to this
angle. For a randomly oriented grain with $z=6$, the closest contact lies
roughly in the range of $\pm \pi/6$ around the principal direction. Let us
imagine that it were possible for a grain to have $z\gg 6$. Now, the contacts
closest to the major direction are spread within a much smaller range $\pm
\pi/z$, so that much narrower widths of $\bar{\bf f}(\phi)$ can be achieved. Of
course, the situations described above are very extreme because in reality all
contacts will be involved in the force equilibrium.  In any case, these simple
examples illustrate that the coordination number introduces a natural scale for
the width due to steric exclusion effects.

If the polydispersity is small, only very few grains will have more
than six neighbours, and the peaks and valleys in $\bar{\bf f}(\phi)$
will therefore have a minimum width of approximately $\pi/3$. This is
why a description using only a few terms in a Fourier expansion works,
as was already clear from existing numerical results.

\section{Frictionless case at arbitrary order}\label{appvandermonde} 
 
In this appendix we analytically solve the linear problem of
(\ref{arbitrary}). Due to the symmetry of $\fnap$ we can write

\begin{equation}
\fnap = 1 + \sum_{k=1}^{N} q_k \, \sin\left( 2k\phi +\frac{(k-1)\pi}{2} \right)~.
\end{equation}
Comparing to (\ref{fn1}), we find $q_k=(-1)^{(k-1)/2} a_k$ for odd $k$, while $q_k=(-1)^{k/2} b_k$ for even $k$. In particular, $q_1 = 2\tau$. The even derivatives can be written as 

\begin{equation}
\frac{\partial^{2l}}{\partial \phi^{2l}}\, \fnap \, = (-1)^{l} \,
\sum_{k=1}^{N} q_k (2k)^{2l} \,\sin\left( 2k\phi +\frac{(k-1)\pi}{2} \right)
~, 
\end{equation}
and since all sine terms evaluated at $\phi=3\pi/4$ yield $-1$, we find for $l\neq0$

\begin{equation}
\left. \frac{\partial^{2l}}{\partial \phi^{2l}}\, \fnap \, 
\right|_{\phi=\frac{3\pi}{4}}= (-1)^{l+1} \, \sum_{k=1}^{N} q_k (2k)^{2l} ~. 
\end{equation}
The linear problem of (\ref{arbitrary}) can now be written in the form of a Vandermonde matrix, 

\begin{eqnarray}\label{vdmonde}
\left( 
\begin{array}{cccc}
1 & 1 & \cdots & 1 \\
x_1 & x_2 & \cdots & x_N \\
x_1^2 & x_2^2 & \cdots & x_N^2 \\
. & & & . \\
. & & & . \\
. & & & . \\
x_1^{N-1} & x_2^{N-1} & \cdots & x_N^{N-1}\\
\end{array}
\right) 
\left(
\begin{array}{c}
q_1 \\
q_2 \\
q_3 \\
. \\
. \\
. \\
q_{N}\\
\end{array}
\right) 
=
\left(
\begin{array}{c}
1 \\
0 \\
0 \\
. \\
. \\
. \\
0 \\
\end{array}
\right)~,
\end{eqnarray}
with $x_k=4k^2$. The inverse $A_{jk}$ of this matrix can be expressed
explicitly in terms of Lagrange polynomials as~\cite{vandermonde}

\begin{equation}
P_j(x) = \prod_{k=1, k \ne j}^N \frac{x-x_k}{x_j-x_k} = \sum_{n=1}^N A_{jn} \, x^{n-1}~.
\end{equation}
We are interested in the solution for $q_1=2\tau$, which for (\ref{vdmonde}) simply reads

\begin{eqnarray}
q_1 &=& A_{11} = P_1(0) = \prod_{k=2}^N \frac{x_k}{x_k-x_1} \nonumber \\
&=& \prod_{k=2}^N \frac{k^2}{(k+1)(k-1)}=\frac{2N}{N+1}~,
\end{eqnarray} 
so that $\tau=N/(N+1)$, (\ref{tauN}). Similarly, the other $q_k$ follow from

\begin{equation}
q_k = \prod_{k=1,k\neq j}^{N} \frac{k^2}{(k+j)(k-j)}~.
\end{equation}

\section{Details for the anisotropic frictionless optimization}
\label{appanis}
We start from the expression for the deviatoric stress in terms of all
parameters and Fourier coefficients, equation (\ref{tauanisfric}),
where we put $c_2=d_1=0$ because the tangential force components are
zero. This reads
\begin{equation}
\tau=\frac{a_1(2-p_1^2)+p_1(2-b_2)}{4}~.\label{tauanis}
\end{equation}
The physical constraints on $a_1$ and $b_2$ are the same as for the
isotropic case discussed in section~\ref{subsecfrictionless}, but the
optimization target is now given by (\ref{tauanis}), instead of
the $\tau=a_1/2$ we had before. The constraints follow from $\fnap$ and its
second derivative being nonnegative at $\phi=3\pi/4$:
\begin{eqnarray}
b_2&\leq&1-a_1(1+p_1/2)\\
b_2&\geq&-a_1/4~.
\end{eqnarray}
The solution to this linear program is then found to be
\begin{eqnarray}
a_1&=&4/(2p_1+3)\\
b_0&=&3/(2p_1+3)\\
b_2&=&-1/(2p_1+3)~,
\end{eqnarray}
which corresponds to
\begin{equation}
\taumax=\frac{7p_1+8}{8p_1+12}~,
\end{equation}
which is the result stated in equation (\ref{taumaxanis}).
 
\end{document}